\documentclass[a4paper,twocolumn,]{emulateapj}

\usepackage{epsfig,wrapfig,color,sidecap}
\usepackage{index}
\usepackage{subfigure}

\usepackage{multirow}

\usepackage{colordvi}

\usepackage[usenames,dvipsnames]{xcolor}
\usepackage[normalem]{ulem}

\usepackage{times}
\usepackage[utf8x]{inputenc}
\usepackage{mathrsfs}

\newcommand {\er}[1] {\textcolor{red}{\sout{#1}}}

\shorttitle{}
\shortauthors{Benyamin et al.}

\RequirePackage[colorlinks=true
,urlcolor=blue
,anchorcolor=blue
,citecolor=blue
,filecolor=blue
,linkcolor=blue
,menucolor=blue
,pagecolor=blue
,linktocpage=true
,pdfproducer=medialab
]{hyperref}

\begin{document}

\bibliographystyle{authordate1}

\title{Lower limits on the nucleosynthezis of $^{44}$T\MakeLowercase{i} and $^{60}$F\MakeLowercase{e} in the dynamic spiral-arms model}

\author{David Benyamin \& Nir J. Shaviv} 
\affil{The Racah Institute of physics, The Hebrew University of Jerusalem, Jerusalem 91904, Israel}

\begin{abstract}

We have previously focused on studying the electron-capture isotopes within the dynamic spiral-arms model and empirically derived the energy dependence of the electron attachment rate using the observation of $^{49}$Ti/$^{49}$V and $^{51}$V/$^{51}$Cr ratios in cosmic rays \citep{EC}. We have also shown how this relation recovers the energy dependence seen in the lab measurements \citep{Letaw}. In this work we use this relation to construct the $^{44}$Ca/$^{44}$Ti ratio and place a lower limit on the amount of $^{44}$Ti that is required to be nucleosynthesized at the source. The results also imply that the acceleration process of the radioisotopes cannot be much longer than a century time scale (or else the required nucleosynthesized amount has to be correspondingly larger). We also provide a similar lower limit on the source $^{60}$Fe by comparing to the recently observed $^{60}$Fe/$^{56}$Fe \citep{Iron60}.

\end{abstract}


\section{introduction}
\label{sec:intro}
\maketitle

In \cite{BoverC,Iron} we developed the first cosmic ray (CRs) propagation code that includes dynamic spiral-arms as the main source of the CRs in the galaxy, and showed how changing the CRs source distribution from the ``standard"  azimuthal symmetry to a dynamic spiral-arms source distribution solves several ``standard" model anomalies.

Within the Iron group nuclei (Scandium through Nickel) there are a few CR isotopes that are known to decay through electron-capture (EC) in the lab, these isotopes can provide an interesting fingerprint on the process of re-acceleration \citep[e.g.,][and references therein]{StrongReview}. In \cite{EC}, we focused on investigating these isotopes and showed that, in principle,  they can also be used to constrain cosmic rays propagation models, though present day uncertainties in the nuclear cross-sections is a limitation. 

Our model considers $^{44}$Ti, $^{49}$V, $^{51}$Cr, $^{53}$Mn, $^{55}$Fe, $^{57}$Co and $^{59}$Ni as EC isotopes\footnote{We note that $^{54}$Mn is also an EC isotope. In our calculations it decay immediately since its $\beta$ decay mode have half life time that is significantly shorter than the typical propagation time.} whose effective half-life can be governed by the electron attachment rate or radioactive decay. The time scale for stripping electrons by the ISM for these isotopes is roughly $\tau_\mathrm{stripping} \approx 5 \times 10^{-3}$\,Myr \citep{Letaw}. For the $^{44}$Ti, $^{49}$V, $^{51}$Cr, $^{55}$Fe and $^{57}$Co, the decay time scale is on the order of several days to a few years, much smaller than $\tau_\mathrm{stripping}$. This implies that we can neglect the stripping process for these isotopes and assume that they decay immediately after they attach an electron from the ISM. However, the EC half life time of $^{53}$Mn and $^{59}$Ni is 3.7\,Myr and 0.076\,Myr respectively, which is much longer than  $\tau_\mathrm{stripping}$. Here one can neglect the decay process and assume that these isotopes will become stripped of their electrons before being able to decay, and can therefore be assumed to be stable.  In \cite{EC}, we considered isotopes governed by the attachment time scale, and empirically obtained the energy dependence of this process using the observation of $^{49}$Ti/$^{49}$V and $^{51}$V/$^{51}$Cr ratios.

When an EC isotope is created through fusion, it has a relatively low energy within the star or subsequent supernova.  This leads to a very high electron attachment cross-section, such that it will decay to its daughter isotope if produced. Several observations detected hard X-ray lines from supernova remnants (SNRs), such as Casssiopeia A and SN1987A, which are associated with the decay of $^{44}$Ti to $^{44}$Sc, 67.9\,KeV and 78.4\,KeV, and the decay of $^{44}$Sc to $^{44}$Ca, 1.157\,MeV (OSSE \citealt{OSSE}, COMPTEL \citealt{COMPTEL}, BeppoSAX \citealt{BeppoSAX} and $\gamma$-rays, \citealt{1987}). Since $^{44}$Ti is an EC isotope with a half life of 60 years, it implies two things. First, the $^{44}$Ti should have been formed within this time scale preceding the supernova. Second, if this $^{44}$Ti is to accelerate and become $^{44}$Ti CRs, it should be accelerated through the SNR shocks, and get stripped, before being able to decay to its daughter isotope.  

$^{44}$Ti is produced through the $^{40}$Ca($\alpha,\gamma$)$^{44}$Ti reaction \citep{The}, which requires a rich $\alpha$ particles supply, as is the case inside a core-collapse (Type II) supernova  during the $\alpha$-rich freeze-out phase. \cite{The} also showed the importance of secondary reactions such as, $^{45}$V(p,$\gamma$)$^{46}$Cr, $^{44}$Ti($\alpha$, p)$^{47}$V and $^{44}$Ti($\alpha,\gamma$)$^{48}$Cr, on the rate of production and the amount of $^{44}$Ti in the supernova explosion, but due to the unstable nature of these isotopes, it is hard to measure the reactions in the lab and provide meaningful constraints.

\cite{Woosley} constrained the production of $^{44}$Ti in SN 1987A using the $^{44}$Ca/$^{56}$Fe ratio of CRs reaching the solar system. Given that $^{44}$Ca is mainly produced by the decay of $^{44}$Ti, they conclude that the $^{44}$Ti/$^{56}$Fe ratio at the source is about the same as the $^{44}$Ca/$^{56}$Fe ratio in CRs reaching the solar system. In later work (\citealt{Diehl}, and references within) this result was recovered and extended for SN 1987A and Cas A.

The connection between $^{60}$Fe and $\gamma$-ray astronomy is extensively discussed in \cite{Diehl2011}. The 1.173\,MeV and 1.332\,MeV lines associated with the decay modes of $^{60}$Fe were detected by the space-based telescopes RHESSI \citep{Smith2004} and INTEGRAL/SPI \citep{Harris2005}, which give an instantaneous snapshot of the on-going nucleosynthesis of this isotope in the Milky Way \citep{Prantzos2010,Diehl2013}.

The production of $^{60}$Fe is associated with core-collapse supernovae, which is expected to be produced in two locations before the supernovae explosion---the neon shell and at the base of the helium shell. In the neon shell, $^{22}$Ne and $^{25,26}$Mg are mixed into the superheated neon burning region, which provide free neutrons to be captured by the $^{58}$Fe seed. The seed itself is previously produced by the \textsl{s}-process during the helium burning phase, and then by the intermediate radioactivity of $^{59}$Fe, to form $^{60}$Fe. At the base of the helium shell, $^{60}$Fe is produced by mild \textsl{r}-process during explosive helium burning \citep{Woosley1995}.

\cite{Meyer} calculated the nucleosynthesis production of the short-lived radioactive isotopes in massive stars, type Ia supernova and neutron star disruption. In massive stars, they predicted that the ratio between $^{60}$Fe to $^{56}$Fe should be around $3 \times 10^{-5}$.

Recently, \cite{Iron60} reported the observation of $^{60}$Fe using the ACE-CRIS instrument in the energy range of 195\,MeV to 500\,MeV. They detected 15 $^{60}$Fe nuclei,  a total Fe number of  $3.55 \times 10^{5}$, and calculated the $^{60}$Fe/$^{56}$Fe and $^{60}$Fe/Fe ratios to be $(4.6 \pm 1.7) \times 10^{-5}$ and $(3.9 \pm 1.4) \times 10^{-5}$ respectively. Using the leaky-box model, they concluded that the ratios at the source are $(7.5 \pm 2.9) \times 10^{-5}$ and $(6.2 \pm 2.4) \times 10^{-5}$ respectively.

We begin in \S\ref{sec:model} by briefly describing the model we developed and our nominal model parameters. In \S\ref{sec:results} we carry out an analysis of the model to find the amount of primary $^{44}$Ti and $^{60}$Fe required to explain the observations obtained by CRIS, using a more modern 3D model than the leaky-box model, namely, the dynamic spiral-arms model. The implications of these results are then discussed in \S\ref{sec:discussion}.

\section{The numerical model}
\label{sec:model}

In \cite{BoverC}, we developed a fully three dimensional numerical code describing the diffusion of CRs in the Milky Way. The code is presently the only model to consider {\em dynamic} spiral arms as the main source of the CR particles. With the model, \cite{BoverC} recovered the B/C ratio and showed how the dynamics of the arms is important for understanding the behavior of nuclei secondaries to primaries ratio, which below 1\,GeV/nuc.\ increase with the energy.

In \cite{Iron} we upgraded the code to be faster and more accurate and showed how a spiral-arms model, unlike a disk-like model, can explain the discrepancy between the grammage required to explain the B/C ratio and the sub-Fe/Fe ratio. The optimal parameters of the model are summarized in table\ \ref{table:parameters}.

\begin{table}[h]
\centering
\caption{Nominal Model Parameters}
\begin{tabular}{ c c c  c}
\hline
Parameter & Definition & Model value\\
\hline
\vspace*{2mm}
$z_\mathrm{h}$ & Half halo height & 250 ~pc\\
$D_0$ & Diffusion coefficient & $1.2 \times 10^{27}$   ~cm$^2$/sec\\
\vspace*{2mm}
 & normalization &\\
 \vspace*{2mm}
$\delta$ & Spectral index & 0.4 
\\
\vspace*{2mm}
$\tau_\mathrm{arm}$ & Last spiral arm passage & 5 ~Myr & \\ 
\vspace*{2mm}
$i_4$ & 4-arms set's pitch angle & 28$^\circ$ & \\
\vspace*{2mm}
$i_2$ & 2-arms set's pitch angle & $11^{\circ}$ & \\
\multirow{2}{*}{$\Omega_4$} & Angular velocity of & \multirow{2}{*}{15 (km/s) kpc$^{-1}$} \\
\vspace*{2mm}
 & the 4-arms set & & \\
\multirow{2}{*}{$\Omega_2$} & Angular velocity of & \multirow{2}{*}{25 (km/s) kpc$^{-1}$} \\
\vspace*{2mm}
 & the 2-arms set & & \\
\multirow{2}{*}{$f_\mathrm{SN,4}$} & Percentage of SN in & \multirow{2}{*}{48.4\%} \\
\vspace*{2mm}
 & the 4-arms set & & \\
\multirow{2}{*}{$f_\mathrm{SN,2}$} & Percentage of SN in & \multirow{2}{*}{24.2\%} \\
\vspace*{2mm}
 & the 2-arms set & & \\
\multirow{2}{*}{$f_\mathrm{SN,CC}$} & Percentage of core collapse & \multirow{2}{*}{8.1\%}\\
\vspace*{2mm}
 & SNe in the disk & & \\
\multirow{2}{*}{$f_\mathrm{SN,Ia}$} & Percentage of  & \multirow{2}{*}{19.3\%}\\
 & SN Type Ia & & \\
\hline
\end{tabular}
\label{table:parameters}
\end{table}

Our code is different from present day simulations (such as {\sc galprop}, \citealt{StrongNucleons}, and {\sc dragon}, \citealt{Dragon}) which solve the diffusion partial differential equations (PDE) in that we are using a Monte Carlo methodology. It allows for more flexibility in adding various physical aspects to the code (such as the spiral arm advection), though at the price of reduced speed. The full details of the code and of the the model are found in \cite{BoverC,Iron}.

In \cite{EC} we focused on the EC isotopes and carried out a full parameter analysis of the electron attachment cross-section formula using measurements of $^{49}$Ti/$^{49}$V and $^{51}$V/$^{51}$Cr ratios in cosmic rays. An empirical relation was derived from these results and is here applied to $^{44}$Ti, $^{49}$V, $^{51}$Cr, $^{55}$Fe and $^{57}$Co isotopes. This relation is $\sigma_{a}(E,Z)=N(z_\mathrm{h}) \times Z^{4.5} \times {(E/\mathrm{500\,MeV})^{-1.8}}$, with a normalization given by $N_{SA}(z_\mathrm{h},\tau_\mathrm{arm})=7.98 \times 10^{-5} ~$mb$ \times (\tau_\mathrm{arm}/10~$Myr$)^{-0.278} \times (z_\mathrm{h}/1~$kpc$)^{0.236}$. The full details on the analysis are found in \cite{EC}.

For $^{53}$Mn and $^{59}$Ni the half life time for the EC decay is 3.7\,Myr and 0.076\,Myr respectively, which is much longer than  $\tau_\mathrm{stripping}\approx 5 \times 10^{-3}$Myr \citep{Letaw}. Consequently,  this allows one to neglect the decay process and assume that these isotopes will become stripped of their electrons before decaying and remain stable. For these isotopes, it is irrelevant to apply the above formula, as their identity will not change.

\section{Results}
\label{sec:results}

\subsection{Primary $^{44}$Ti}
\label{sec:Ti44}

\begin{figure}[ht]
\centerline{\includegraphics[width=3.5in]{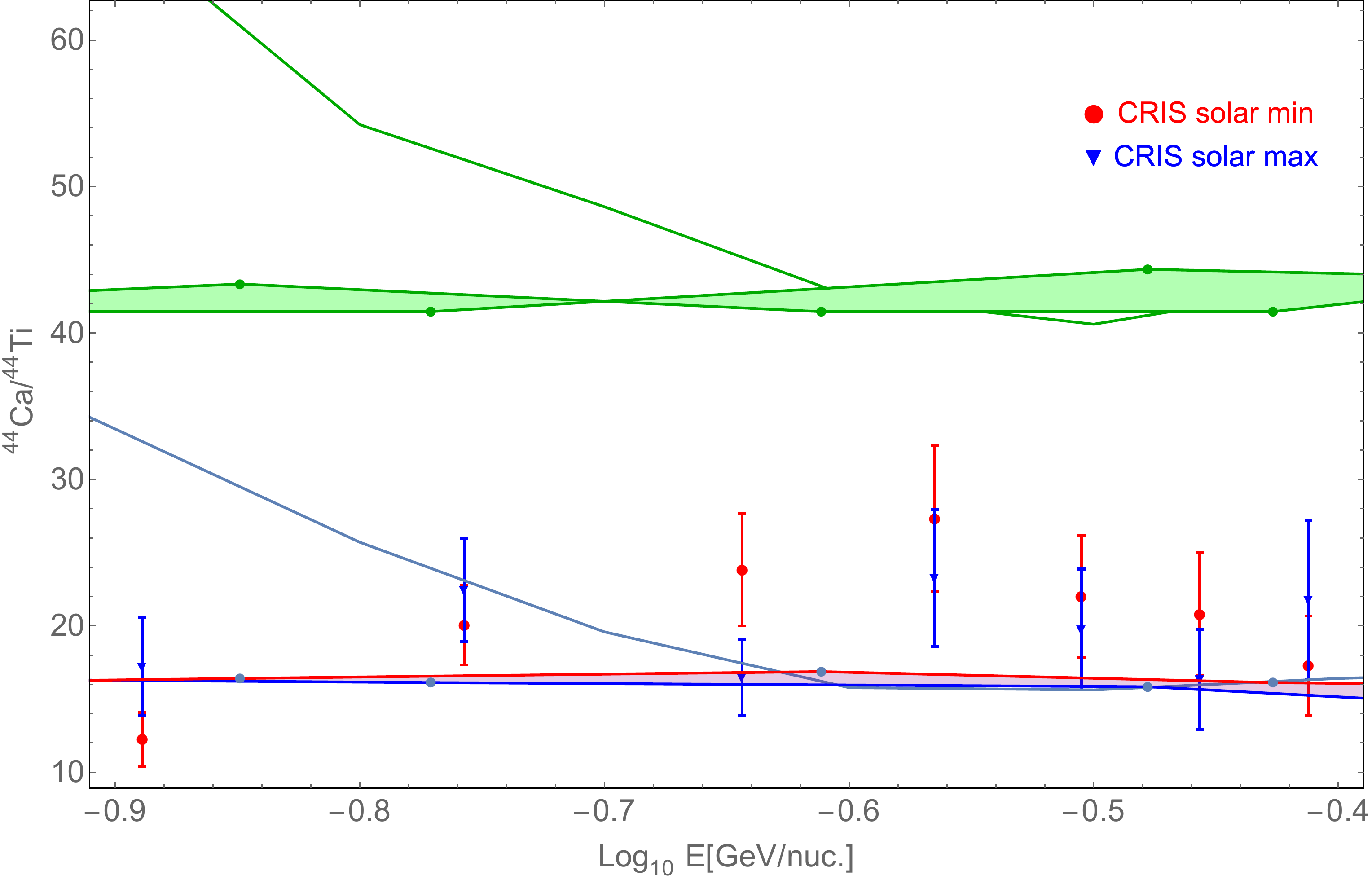} }
\caption{The $^{44}$Ca/$^{44}$Ti ratio, the green lines represent the simulation which we did not include any primary $^{44}$Ti. It can be seen that the simulation is higher by a factor of 2 from the observations (ACE/CRIS, \citealt{scott}). The shaded area is the correction to the simulation due to solar modulation (taking for solar minimum, $\phi=513\,MV$ and for solar maximum, $\phi=923\,MV$). The blue lines are obtained after adding an amount of $^{44}$Ti/Fe$=0.4\%$ to the initial composition.
\label{fig:44}}
\end{figure}

We begin by implementing the attachment rate formula to the EC isotopes, and specifically to $^{44}$Ti. This allows us to predict the amount of $^{44}$Ti and compare it its daughter isotope, $^{44}$Ca. The results are depicted in fig.\ \ref{fig:44}. With our simulation, we find a ratio that is higher by about a factor of 2 from the observations. This can be explained by the fact that we did not include any $^{44}$Ti in the initial composition---any additional $^{44}$Ti that is initially present will decrease the $^{44}$Ca/$^{44}$Ti  ratio. In order for the $^{44}$Ti to not decay, it has to quickly accelerate by the SNR shocks to a sufficiently high energy and be stripped of its electrons, compared with its decay half life of 60 years. By fitting our model results to the observations, we can determine the minimal amount of $^{44}$Ti in the initial composition which escape the SNR obtained if the acceleration is fast. If some of the $^{44}$Ti can decay then the required $^{44}$Ti at the source should be correspondingly higher.  

The optimal amount of primary $^{44}$Ti required to recover \cite{scott}'s observations is $^{44}$Ti/Fe$=0.40\% \pm 0.03\%$, which means that the ratio $^{44}$Ti/$^{56}$Fe is $=0.44\% \pm 0.03\%$.

\cite{scott} also report the observations of the $^{44}$Ca/$^{56}$Fe which is about $0.5\% \pm 0.1\%$. According to \cite{Diehl} and \cite{Woosley} the initial $^{44}$Ti/$^{56}$Fe ratio should be about the same as the $^{44}$Ca/$^{56}$Fe ratio measured in CRs reaching the solar system, which is in good agreement with our results.

\newpage
\subsection{Primary $^{60}$Fe}
\label{sec:Iron60}

The next step is to estimate \er{is} the amount of $^{60}$Fe in the initial composition. To do so, we carry out a similar analysis to the one described above for $^{44}$Ti, and estimate the initial amount of $^{60}$Fe required to fit the recent CRIS results \citep{Iron60}.

\begin{figure}[ht]
\centerline{\includegraphics[width=3.5in]{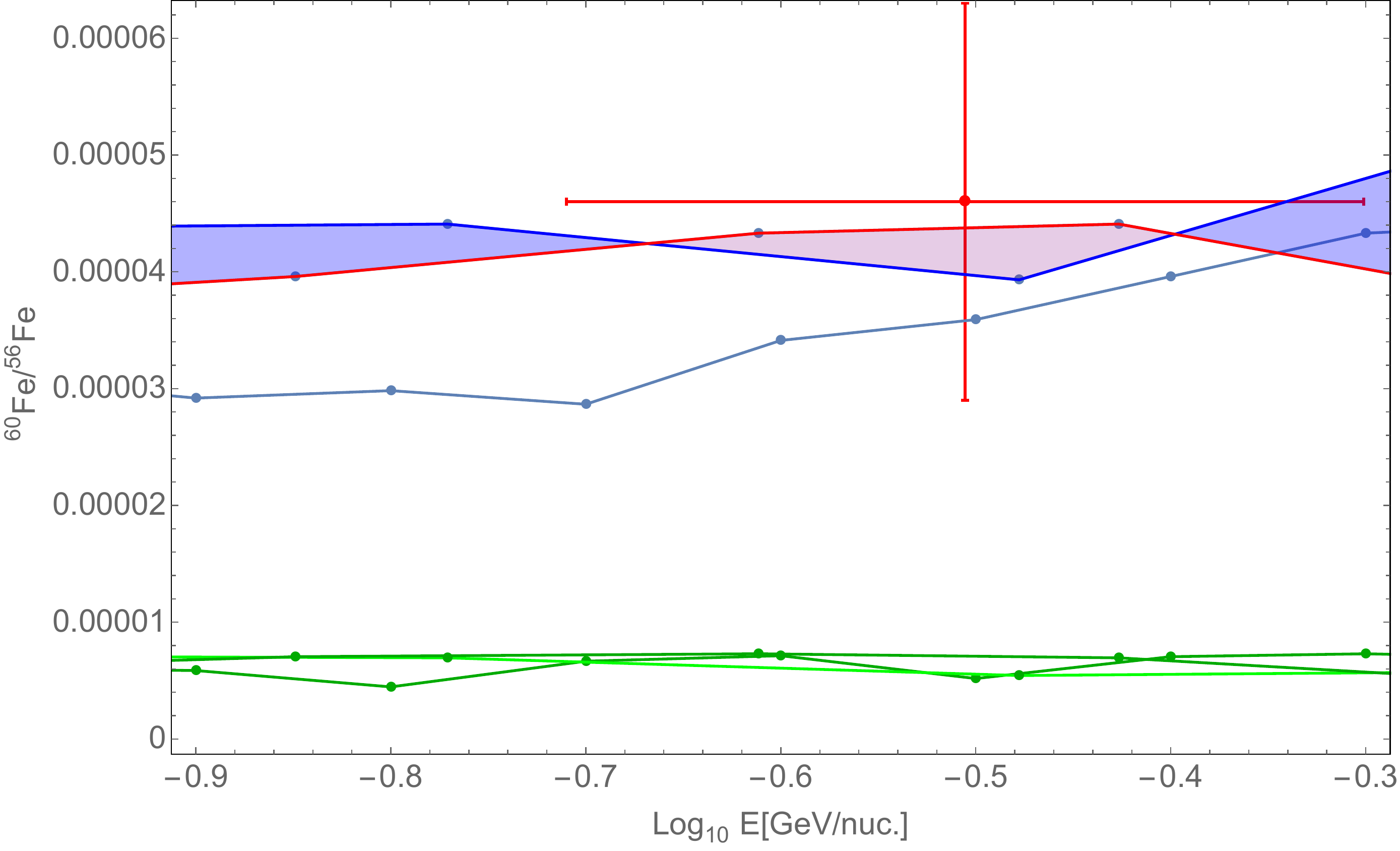} }
\caption{Same as fig.\ \ref{fig:44}. The green lines correspond to the $^{60}$Fe/$^{56}$Fe ratio obtained without the primary component. The shaded area is the correction to the simulation after solar modulation is added. The blue lines are derived after an amount of $^{60}$Fe/$^{56}$Fe$=(4.5 \pm 2) \times 10^{-5}$ is added to the initial composition. The Data is of the CRIS experiment \citep{Iron60}.}
\label{fig:60}
\end{figure}

Fig.\ \ref{fig:60} depicts the $^{60}$Fe/$^{56}$Fe ratio in our model, with and without the primary $^{60}$Fe. The optimal fit corresponds an to initial $^{60}$Fe/$^{56}$Fe ratio of $(4.5 \pm 2) \times 10^{-5}$. Our results agree with \cite{Meyer} who predict $^{60}$Fe/$^{56}$Fe=$3 \times 10^{-5}$ and with \cite{Iron60}'s estimate of $^{60}$Fe/$^{56}$Fe=$(7.5 \pm 2.9) \times 10^{-5}$. \footnote{{Although the spiral-arms model agrees with the predictions of \cite{Iron60}, we carried out the same analysis with a disk-like model as well (using the same estimates as in \citealt{Iron60}). We found in this case that one requires a $^{60}$Fe/$^{56}$Fe ratio of $(6 \pm 2.5) \times 10^{-5}$, which is closer to the predictions of \cite{Iron60}.}}

\newpage
\section{Discussion \& Summary}
\label{sec:discussion}

It is generally accepted that the bulk of the galactic cosmic rays (whether in number or energy) are accelerated in supernova remnants \citep[e.g.,][and references therein]{FermiSNR}, while the production of Iron group nuclei is through fusion in the last evolutionary phase of the progenitor stars or the SN event itself.  Indeed, $^{44}$Ti is spectrally detected in SNRs Casssiopeia A and SNR 1987A \citep{OSSE, COMPTEL, BeppoSAX, 1987}. Since $^{44}$Ti is produced through the reaction $^{40}$Ca($\alpha,\gamma$)$^{44}$Ti \citep{The}, its detection in remnants is evidence of a rich $\alpha$ particle supply in the supernova explosion, which existed inside the core-collapse supernova during the $\alpha$-rich freeze-out phase. However, if the acceleration process of the CR isotopes coming from the SNe is relatively long, then by the time the nuclei are accelerated those nuclei which are unstable through electron-capture should decay. A short acceleration process will however strip the nuclei of their electrons and allow them to be long lived cosmic rays. 

There is however another source of $^{44}$Ti in the cosmic rays---CR nuclei are also created through spallation during their propagation in the galaxy. Since they are formed stripped, these EC unstable isotopes can survive as long as they remain at high energies. As a consequence, nuclei which decay through EC, have a mean half-life time which depends strongly on the energy. This can be seen with the \cite{ACE} measurements showing how the $^{49}$Ti/$^{49}$V and $^{51}$V/$^{51}$Cr ratios decrease with energy, as expected from the longer decay time of the EC isotopes, $^{49}$V and $^{51}$Cr, at higher energies.

In our previous analyses \citep{BoverC,Iron} we showed how our propagation model can be used to  describe the cosmic ray propagation by fitting the secondary to primary ratios in the Beryllium-Oxygen and Scandium-Nickel elements groups.

\cite{Jones} and \cite{Niebur} suggested that a standard diffusion model cannot explain the behaviour of EC isotopes and cannot explain the decrease in the ratios of the daughter EC isotopes to the EC isotopes, for example, the ratios $^{49}$Ti/$^{49}$V and $^{51}$V/$^{51}$Cr. \cite{Jones} and \cite{Niebur} were on agreement that nominal diffusion models cannot give a strong enough decrease as the energy increases. Their solution for the decrease of $^{49}$Ti/$^{49}$V and $^{51}$V/$^{51}$Cr was to add to their propagation model an ad hoc assumption on the reacceleration of the nuclei on their way to earth, in order to fit these observations.

In our previous work on EC isotopes \citep{EC}, we suggested another explanation for the decrease in the ratio of $^{49}$Ti/$^{49}$V and $^{51}$V/$^{51}$Cr. We showed that a energy dependent  cross-section for the attachment of electrons from the ISM can explain the observed behavior, without having to add any additional primary cosmic rays at the source. When the isotope attaches an electron, it subsequently decays through EC. The fitted functional form for the electron attachment cross-section that we obtained in \cite{EC} is $\sigma_{a}(E,Z)=N(z_\mathrm{h}) \times Z^{4.5} \times {(E/\mathrm{500\,MeV})^{-1.8}}$, with a normalization given by $N_{SA}(z_\mathrm{h},\tau_\mathrm{arm})=7.98 \times 10^{-5} ~$mb$ \times (\tau_\mathrm{arm}/10~$Myr$)^{-0.278} \times (z_\mathrm{h}/1~$kpc$)^{0.236}$.

With the help of the empirical fit obtained in \cite{EC}, we simulated here the $^{44}$Ca/$^{44}$Ti ratio and found that the ratio is higher than the observations by a factor of about 2. This can be explained away by adding $^{44}$Ti to the list of injected isotopes, as is corroborated with the observations \citep{OSSE, COMPTEL, BeppoSAX, 1987}. We found out that the amount of $^{44}$Ti/$^{56}$Fe required to be injected as part of the initial composition is $0.44\% \pm 0.03\%$ in order to match the CRIS observations \citep{scott}. Our results agree with \cite{Diehl} and \cite{Woosley} who predict it to be about the same as the $^{44}$Ca/$^{56}$Fe ratio measured in CRs reaching the solar system, which is about $0.5\% \pm 0.1\%$ \citep{scott}.

Recently, \cite{Iron60} reported the detection and measurement of $^{60}$Fe in cosmic rays using the ACE-CRIS instrument. The ratios $^{60}$Fe/$^{56}$Fe and $^{60}$Fe/Fe found are $(4.6 \pm 1.7) \times 10^{-5}$ and $(3.9 \pm 1.4) \times 10^{-5}$ respectively. We found that we need to have an initial $^{60}$Fe/$^{56}$Fe ratio of $(4.5 \pm 2) \times 10^{-5}$ to the initial composition in order to fit the observed $^{60}$Fe/$^{56}$Fe. Our results also agree with \cite{Meyer} who predict $^{60}$Fe/$^{56}$Fe=$3 \times 10^{-5}$ and with \cite{Iron60} who estimated a ratio of $^{60}$Fe/$^{56}$Fe=$(7.5 \pm 2.9) \times 10^{-5}$.

As a word of caution, one should emphasize that some of the EC radioactive isotopes could decay during the acceleration phase before escaping the SNR, thus, the amount of $^{44}$Ti/$^{56}$Fe $=0.44\% \pm 0.03\%$ and of $^{60}$Fe/$^{56}$Fe$=(4.5 \pm 2) \times 10^{-5}$ which one requires to add to the initial composition of cosmic rays is actually only a lower limit on the nucleosynthesis of these isotopes. 

\section*{Acknowledgements} 

The authors wish to thank Michael Paul for use valuable suggestions. NJS gratefully acknowledges the support of the Israel Science Foundation (grant no. 1423/15) and the I-CORE Program of the Planning and Budgeting Committee and the Israel Science Foundation (center 1829/12). 

\def\jcap{J.\ Cos.\ Astropart.\ Phys.}
\def\na{N.\ Astron.}

\bibliography{Ti44}

\begin{thebibliography}{}

\bibitem[\protect\citename{{Ackermann}, }2013]{FermiSNR}
{Ackermann}, M. et~al. 2013.
\newblock {Detection of the Characteristic Pion-Decay Signature in Supernova
  Remnants}.
\newblock {\em Science}, {\bf 339}(Feb.), 807--811.

\bibitem[\protect\citename{{Benyamin} {\em et~al.\ }\relax, }2014]{BoverC}
{Benyamin}, D., {Nakar}, E., {Piran}, T., \& {Shaviv}, N.~J. 2014.
\newblock {Recovering the Observed B/C Ratio in a Dynamic Spiral-armed Cosmic
  Ray Model}.
\newblock {\em \apj}, {\bf 782}(Feb.), 34.

\bibitem[\protect\citename{{Benyamin} {\em et~al.\ }\relax, }2016]{Iron}
{Benyamin}, D., {Nakar}, E., {Piran}, T., \& {Shaviv}, N.~J. 2016.
\newblock {The B/C and Sub-iron/Iron Cosmic Ray Ratios--Further Evidence in
  Favor of the Spiral-Arm Diffusion Model}.
\newblock {\em \apj}, {\bf 826}(July), 47.

\bibitem[\protect\citename{{Benyamin} {\em et~al.\ }\relax, }2017]{EC}
{Benyamin}, D., {Shaviv}, N.~J., \& {Piran}, T. 2017.
\newblock {\em Can Electron-capture isotopes constrain spiral arms cosmic-ray
  propagation models?}
\newblock submitted to \apjl.

\bibitem[\protect\citename{{Binns} {\em et~al.\ }\relax, }2016]{Iron60}
{Binns}, W.~R., {Israel}, M.~H., {Christian}, E.~R., {Cummings}, A.~C., {de
  Nolfo}, G.~A., {Lave}, K.~A., {Leske}, R.~A., {Mewaldt}, R.~A., {Stone},
  E.~C., {von Rosenvinge}, T.~T., \& {Wiedenbeck}, M.~E. 2016.
\newblock {Observation of the $^{60}$Fe nucleosynthesis-clock isotope in
  galactic cosmic rays}.
\newblock {\em Science}, {\bf 352}(May), 677--680.

\bibitem[\protect\citename{{di Bernardo} {\em et~al.\ }\relax, }2010]{Dragon}
{di Bernardo}, G., {Evoli}, C., {Gaggero}, D., {Grasso}, D., \& {Maccione}, L.
  2010.
\newblock {Unified interpretation of cosmic ray nuclei and antiproton recent
  measurements}.
\newblock {\em Astroparticle Physics}, {\bf 34}(Dec.), 274--283.

\bibitem[\protect\citename{{Diehl}, }2013]{Diehl2013}
{Diehl}, R. 2013.
\newblock {Nuclear astrophysics lessons from INTEGRAL}.
\newblock {\em Reports on Progress in Physics}, {\bf 76}(2), 026301.

\bibitem[\protect\citename{{Diehl} {\em et~al.\ }\relax, }2006]{Diehl}
{Diehl}, R., {Prantzos}, N., \& {von Ballmoos}, P. 2006.
\newblock {Astrophysical constraints from gamma-ray spectroscopy}.
\newblock {\em Nuclear Physics A}, {\bf 777}(Oct.), 70--97.

\bibitem[\protect\citename{{Diehl} {\em et~al.\ }\relax, }2011]{Diehl2011}
{Diehl}, R., {Hartmann}, D.~H., \& {Prantzos}, N. (eds). 2011.
\newblock {\em {Astronomy with Radioactivities}}.
\newblock Lecture Notes in Physics, Berlin Springer Verlag, vol. 812.

\bibitem[\protect\citename{{Grebenev} {\em et~al.\ }\relax, }2012]{1987}
{Grebenev}, S.~A., {Lutovinov}, A.~A., {Tsygankov}, S.~S., \& {Winkler}, C.
  2012.
\newblock {Hard-X-ray emission lines from the decay of $^{44}$Ti in the remnant
  of supernova 1987A}.
\newblock {\em \nat}, {\bf 490}(Oct.), 373--375.

\bibitem[\protect\citename{{Harris} {\em et~al.\ }\relax, }2005]{Harris2005}
{Harris}, M.~J., {Kn{\"o}dlseder}, J., {Jean}, P., {Cisana}, E., {Diehl}, R.,
  {Lichti}, G.~G., {Roques}, J.-P., {Schanne}, S., \& {Weidenspointner}, G.
  2005.
\newblock {Detection of {$\gamma$}-ray lines from interstellar $^{60}$Fe by the
  high resolution spectrometer SPI}.
\newblock {\em \aap}, {\bf 433}(Apr.), L49--L52.

\bibitem[\protect\citename{{Iyudin} {\em et~al.\ }\relax, }1994]{COMPTEL}
{Iyudin}, A.~F., {Diehl}, R., {Bloemen}, H., {Hermsen}, W., {Lichti}, G.~G.,
  {Morris}, D., {Ryan}, J., {Schoenfelder}, V., {Steinle}, H., {Varendorff},
  M., {de Vries}, C., \& {Winkler}, C. 1994.
\newblock {COMPTEL observations of Ti-44 gamma-ray line emission from CAS A}.
\newblock {\em \aap}, {\bf 284}(Apr.), L1--L4.

\bibitem[\protect\citename{{Jones} {\em et~al.\ }\relax, }2001]{Jones}
{Jones}, F.~C., {Lukasiak}, A., {Ptuskin}, V.~S., \& {Webber}, W.~R. 2001.
\newblock {K-Capture cosmic ray secondaries and reacceleration}.
\newblock {\em International Cosmic Ray Conference}, {\bf 5}(Aug.), 1844.

\bibitem[\protect\citename{{Letaw} {\em et~al.\ }\relax, }1985]{Letaw}
{Letaw}, J.~R., {Adams}, Jr., J.~H., {Silberberg}, R., \& {Tsao}, C.~H. 1985.
\newblock {Electron capture decay of cosmic rays}.
\newblock {\em \apss}, {\bf 114}(Sept.), 365--379.

\bibitem[\protect\citename{{Meyer} \& {Clayton}, }2000]{Meyer}
{Meyer}, B.~S., \& {Clayton}, D.~D. 2000.
\newblock {Short-Lived Radioactivities and the Birth of the sun}.
\newblock {\em \ssr}, {\bf 92}(Apr.), 133--152.

\bibitem[\protect\citename{{Niebur} {\em et~al.\ }\relax, }2000]{ACE}
{Niebur}, S.~M., {Binns}, W.~R., {Christian}, E.~R., {Cummings}, A.~C.,
  {George}, J.~S., {Hink}, P.~L., {Israel}, M.~H., {Klarmann}, J., {Leske},
  R.~A., {Lijowski}, M., {Mewaldt}, R.~A., {Stone}, E.~C., {von Rosenvinge},
  T.~T., {Wiedenbeck}, M.~E., \& {Yanasak}, N.~E. 2000 (Sept.).
\newblock {Secondary electron-capture-decay isotopes and implications for the
  propagation of galactic cosmic rays}.
\newblock {\em Pages  406--409 of:} {Mewaldt}, R.~A., {Jokipii}, J.~R., {Lee},
  M.~A., {M{\"o}bius}, E., \& {Zurbuchen}, T.~H. (eds), {\em Acceleration and
  Transport of Energetic Particles Observed in the Heliosphere}.
\newblock American Institute of Physics Conference Series, vol. 528.

\bibitem[\protect\citename{{Niebur} {\em et~al.\ }\relax, }2001]{Niebur}
{Niebur}, S.~M., {Binns}, W.~R., {Christian}, E.~R., {Cummings}, A.~C., {de
  Nolfo}, G.~A., {George}, J.~S., {Hink}, P.~L., {Israel}, M.~H., {Leske},
  R.~A., {Mewaldt}, R.~A., {Stone}, E.~C., {von Rosenvinge}, T.~T.,
  {Wiedenbeck}, M.~E., \& {Yanasak}, N.~E. 2001.
\newblock {CRIS measurements of electron-capture decay isotopes: 37Ar, 44Ti,
  49V, 51Cr, 55Fe, and 57Co}.
\newblock {\em International Cosmic Ray Conference}, {\bf 5}(Aug.), 1675.

\bibitem[\protect\citename{{Prantzos}, }2010]{Prantzos2010}
{Prantzos}, N. 2010.
\newblock {Nucleosynthesis and gamma-ray lines}.
\newblock {\em Page ~18 of:} {\em Eighth Integral Workshop. The Restless
  Gamma-ray Universe (INTEGRAL 2010)}.

\bibitem[\protect\citename{{Scott}, }2005]{scott}
{Scott}, L.~M. 2005.
\newblock {\em {Cosmic-ray energy loss in the heliosphere and interstellar
  reacceleration}}.
\newblock Ph.D. thesis, Washington University, Missouri, USA.

\bibitem[\protect\citename{{Smith}, }2004]{Smith2004}
{Smith}, D.~M. 2004 (Oct.).
\newblock {Gamma-Ray Line Observations with RHESSI}.
\newblock {\em Page ~45 of:} {Schoenfelder}, V., {Lichti}, G., \& {Winkler}, C.
  (eds), {\em 5th INTEGRAL Workshop on the INTEGRAL Universe}.
\newblock ESA Special Publication, vol. 552.

\bibitem[\protect\citename{{Strong} \& {Moskalenko}, }1998]{StrongNucleons}
{Strong}, A.~W., \& {Moskalenko}, I.~V. 1998.
\newblock {Propagation of Cosmic-Ray Nucleons in the Galaxy}.
\newblock {\em \apj}, {\bf 509}(Dec.), 212--228.

\bibitem[\protect\citename{{Strong} {\em et~al.\ }\relax, }2007]{StrongReview}
{Strong}, A.~W., {Moskalenko}, I.~V., \& {Ptuskin}, V.~S. 2007.
\newblock {Cosmic-Ray Propagation and Interactions in the Galaxy}.
\newblock {\em Annual Review of Nuclear and Particle Science}, {\bf 57}(Nov.),
  285--327.

\bibitem[\protect\citename{{The} {\em et~al.\ }\relax, }1996]{OSSE}
{The}, L.-S., {Leising}, M.~D., {Kurfess}, J.~D., {Johnson}, W.~N., {Hartmann},
  D.~H., {Gehrels}, N., {Grove}, J.~E., \& {Purcell}, W.~R. 1996.
\newblock {CGRO/OSSE observations of the Cassiopeia A SNR.}
\newblock {\em \aaps}, {\bf 120}(Dec.), 357--360.

\bibitem[\protect\citename{{The} {\em et~al.\ }\relax, }1998]{The}
{The}, L.-S., {Clayton}, D.~D., {Jin}, L., \& {Meyer}, B.~S. 1998.
\newblock {Nuclear Reactions Governing the Nucleosynthesis of $^{44}$Ti}.
\newblock {\em \apj}, {\bf 504}(Sept.), 500--515.

\bibitem[\protect\citename{{Vink} {\em et~al.\ }\relax, }2001]{BeppoSAX}
{Vink}, J., {Laming}, J.~M., {Kaastra}, J.~S., {Bleeker}, J.~A.~M., {Bloemen},
  H., \& {Oberlack}, U. 2001.
\newblock {Detection of the 67.9 and 78.4 keV Lines Associated with the
  Radioactive Decay of $^{44}$Ti in Cassiopeia A}.
\newblock {\em \apjl}, {\bf 560}(Oct.), L79--L82.

\bibitem[\protect\citename{{Woosley} \& {Hoffman}, }1991]{Woosley}
{Woosley}, S.~E., \& {Hoffman}, R.~D. 1991.
\newblock {Co-57 and Ti-44 production in SN 1987A}.
\newblock {\em \apjl}, {\bf 368}(Feb.), L31--L34.

\bibitem[\protect\citename{{Woosley} \& {Weaver}, }1995]{Woosley1995}
{Woosley}, S.~E., \& {Weaver}, T.~A. 1995.
\newblock {The Evolution and Explosion of Massive Stars. II. Explosive
  Hydrodynamics and Nucleosynthesis}.
\newblock {\em \apjs}, {\bf 101}(Nov.), 181.

\end{thebibliography}

\end{document}